\documentclass[doublecol]{epl2}

\title{Normal heat conductivity in two-dimensional scalar lattices}

\author{A. V. Savin\inst{1} \and V. Zolotarevskiy\inst{2}
\and O. V. Gendelman\inst{2}}
\shortauthor{A. V. Savin\etal}

\institute{
  \inst{1} Semenov Institute of Chemical Physics, Russian Academy of Sciences, Moscow 119991, Russia\\
  \inst{2} Faculty of Mechanical Engineering, Technion -- Israel Institute of Technology,
           Haifa 32000, Israel
}

\abstract{
The paper revisits recent counterintuitive results on divergence of heat conduction coefficient
in two-dimensional lattices. It was reported that in certain lattices with on-site potential,
for which one-dimensional chain has convergent conductivity, for the 2D case it turns out
to diverge. We demonstrate that this conclusion is an artifact caused by insufficient size
of the simulated system. To overcome computational restrictions, a ribbon of relatively
small width is simulated instead of the square specimen. It is further demonstrated that
the heat conduction coefficient in the "long" direction of the ribbon ceases to depend
on the width, as the latter achieves only 10 to 20 interparticle distances. So, one can consider
the dynamics of much longer systems, than in the traditional setting, and still can gain a reliable
information regarding the 2D lattice. It turns out that for all considered models, for which
the conductivity is convergent in the 1D case, it is indeed convergent in the 2D case.
In the same time, however, the length of the system, necessary to reveal the convergence
in the 2D case, may be much bigger than in its 1D counterpart.}

\begin{document}

\maketitle

\section{Introduction}
Heat conductivity in one-dimensional (1D) lattices is a
well-known classical problem related to the microscopic
foundation of Fourier's law. The problem has been first posed in
famous work of Fermi, Pasta, and Ulam (FPU) \cite{FPU}, where
a lack of thermalisation in weakly nonlinear atomic chain has been observed in the first time.
Large number of works has been devoted to this problem, see e.g. the reports
\cite{LLP03,LLP08,Dhar08}.

Modelling of heat conductivity in high-dimensional lattices (2D or 3D)
requires  significant computational resources. Due to this
fact many problems remain unsolved.

The first works on modelling of heat transport in 2D lattices \cite{payton67,mount83}
{\it a priori} suggested finite heat conductivity. One of the first papers to study system
size dependence was probably that by Jackson and Mistriotis \cite{jack89}.
If the heat conductivity of one-dimensional lattice diverges, one may expect a divergence
in a corresponding 2D lattice. Modelling of heat transport in 2D scalar Fermi-Pasta-Ulam lattice \cite{gras02}
shows that the heat conductivity, like in 1D lattice, diverges as a power law of the distance:
$\kappa\sim L^\alpha$. Other studies \cite{Lippi00,gras06,Li12} reveal logarithmic divergence
of the heat conduction coefficient in the 2D case, when in the 1D counterpart it diverges
according to the power law.

One could intuitively expect, that if in the 1D chain the heat conductivity converges, in the 2D
counterpart it will be definitely convergent. This idea was questioned
in works \cite{barik06,barik07}; there the author concluded that the 2D models demonstrate
logarithmic divergence even for convergent 1D version. In report \cite{Dhar08} it was suggested
that these conclusions might be caused by the insufficient system size. We demonstrate
in this paper for a set of typical models, that this latter suggestion is correct, at least as 2D scalar models are considered.
Still, there is a significant difference
between the 1D and 2D cases. In the 2D lattice the convergence of the conductivity occurs
on significantly longer system sizes than in 1D lattices. Possible reason is that in the 1D lattices
phonon scattering on localized nonlinear excitations ensures the convergence of heat conductivity.
In two dimensions the phonons obtain an ability to bypass such sites, so in 2D
lattices the intensity of scattering occurrences decreases relatively to the 1D counterparts.

For this analysis we choose the systems with on-site potential similar to used in \cite{barik06,barik07},
for which the convergence in 1D case is more or less firmly established. Such finite conductivity
was observed, for instance, in the sinh-Gordon \cite{TBSZ} chain and in 1D Frenkel-Kontorova
lattice (discrete sin-Gordon model) \cite{HLZ98,SG03}.  Other example is a chain of rotators,
which possesses the same property \cite{giardina00,GS00}. Recently it was suggested that the class
of 1D systems with convergent heat conductivity is much wider, and includes a broad variety of
dissociating atomic chains \cite{SK14,GS14}. Still, as this issue is still debated, only the
models which cause no objections were selected for current study.

\section{Model}
Let us consider a two-dimensional scalar lattice with dimensionless
Hamiltonian
\begin{eqnarray}
H&=&\sum_{n=-\infty}^{+\infty}\sum_{m=-\infty}^{+\infty}\frac12\dot{u}^2_{n,m}
    +V(u_{n+1,m}-u_{n,m})+\nonumber\\
&& +V(u_{n,m+1}-u_{n,m})+U(u_{n,m}), \label{f1}
\end{eqnarray}
where $u_{n,m}$ -- displacement  of particle $(n,m)$ in the lattice,
$V(\rho)$ -- interaction potential of the nearest neighbor particle,
$U(u)$ -- interaction potential of each particle with a substrate of the lattice.

If all the particles with the second index $m$ displace equally, i.e.
if $u_{n,m}\equiv u_n$ for $m=0,\pm1,\pm2,...$, then the Hamiltonian (\ref{f1})
will take the form of the Hamiltonian of the one-dimensional chain
\begin{equation}
H=\sum_{n=-\infty}^{+\infty}\frac12\dot{u}^2_{n}+V(u_{n+1}-u_{n})+U(u_n). \label{f2}
\end{equation}

Let us consider three typical sets of interaction potentials for which
the 1D lattice (\ref{f2}) has finite heat conductivity.

The first simulated system corresponds to the sinh-Gordon model \cite{TBSZ}
\begin{equation}
V(\rho)=\rho^2/2,~~U(u)=\omega_0[\cosh(u)-1]. \label{f3}
\end{equation}
The second system is a well-known Frenkel-Kontorova model \cite{HLZ98,SG03}
\begin{equation}
V(\rho)=\rho^2/2,~~U(u)=\epsilon[1-\cos(u)], \label{f4}
\end{equation}
and the third considered system is the chain of rotators \cite{giardina00,GS00}
\begin{equation}
V(\rho)=1-\cos(\rho),~~U(u)\equiv 0. \label{f5}
\end{equation}

\section{Modelling of heat conductivity}

Simulation of 2D lattices produces contradictory results due to enormous computational
difficulties. It is well-known that in 1D systems one may require up to $10^4$ particles
in the chain to obtain conclusive evidence on convergence of the heat conduction coefficient \cite{GS14}.
In the 2D counterpart, it is reasonable to require about $10^8$ particles -- far beyond modern
computational capabilities. To overcome this difficulty, we invoke recent results on transition
between one-dimensional and two-dimensional character of the heat transport in channels with
varying width \cite{ZSG}. It was demonstrated that the transition occurs for relatively narrow
channel width, and further broadening of the channel has almost no effect on the character of
the heat transport. So, it is possible to conjecture that even relatively narrow ribbon of the
simulated specimen will exhibit all significant features of complete 2D system. If it is the case,
it is possible to consider long and relatively narrow ribbons, and to obtain conclusive results
with existing computational capabilities. 3D specimen with large aspect ratio was used to obtain
evidence on convergence of the heat conductivity in the FPU lattice \cite{Saito},
but no systematic exploration of transition from 1D to 3D was presented.
Let us consider heat transport in finite 2D stripe
 $1\le n\le N_x$, $1\le m\le N_y$
with periodic boundary conditions for transversal direction $m$, i.e. heat transfer in a
stripe glued into a cylinder. This type of the boundary conditions seems to be most reasonable
to avoid boundary effects. The Hamiltonian  of this system has a form
\begin{eqnarray}
H &=& \sum_{m=1}^{N_y}\left\{\sum_{n=1}^{N_x}\left[\frac12\dot{u}^2_{n,m}
         +V(u_{n,m+1}-u_{n,m})+U(u_{n,m})\right]\right. \nonumber\\
&&\left. +\sum_{n=1}^{N_x-1}V(u_{n+1,m}-u_{n,m})\right\}, \label{f6}
\end{eqnarray}
where $m+1=1$ with $m=N_y$.

For modelling of the heat transport the ends of the cylinder are coupled to Langevin thermostats
with temperatures $T_+$ and $T_-$. The corresponding system of equations of motions reads
\begin{eqnarray}
\ddot{u}_{n,m}&=&-\frac{\partial H}{\partial u_{n,m}}-\gamma\dot{u}_{n,m}+\xi_{n,m},\nonumber\\
&& n=2,...,N_t;\nonumber\\
\ddot{u}_{n,m}&=&-\frac{\partial H}{\partial u_{n,m}},\label{f7}\\
&& n=N_t+1,...,N_t+N_x;\nonumber\\
\ddot{u}_{n,m}&=&-\frac{\partial H}{\partial u_{n,m}}-\gamma\dot{u}_{n,m}+\eta_{n,m},\nonumber\\
&& n=N_t+N_x+1,...,N_t+N_x+N_t-1;\nonumber
\end{eqnarray}
where $m=1,...,N_y$, $u_{1,m}\equiv 0$, $u_{N_t+N_y+N_t,m}\equiv 0$ (fixed ends condition),
$\gamma$ -- relaxation coefficient,  and $\xi_{n,m}$, $\eta_{n,m}$ -- Gaussian white noise,
which models the interaction with the thermostats and is normalized by the following conditions:
\begin{eqnarray}
\langle\xi_{n,m}(t)\rangle=\langle\eta_{n,m}(t)\rangle=\langle\xi_{n,m}(t_1)\eta_{k,l}(t_2)\rangle=0,\nonumber\\
\langle\xi_{n,m}(t_1)\xi_{k,l}(t_2)\rangle=2\gamma T_+\delta_{nk}\delta_{ml}\delta(t_2-t_1),
\nonumber\\
\langle\eta_{n,m}(t_1)\eta_{k,l}(t_2)\rangle=2\gamma T_-\delta_{nk}\delta_{ml}\delta(t_2-t_1).
\nonumber
\end{eqnarray}

System of Eqs. (\ref{f7}) was integrated numerically, using  values
$N_t=20$, $\gamma=0.05$ (time of relaxation $t_r=1/\gamma=20$), $N_x=10$, 20, 40, ..., $N_y=1$, 2, 3,....
As the stationary heat flux is established in the system, one observes
an average temperature gradient along the cylinder axis:
$$
T_n=\frac{1}{N_y}\sum_{m=1}^{N_y}\langle\dot{u}^2_{nm}(t)\rangle_t=
\frac{1}{N_y}\sum_{m=1}^{N_y}\lim_{t\rightarrow\infty}\frac1t\int_0^t\dot{u}^2_{nm}(\tau)d\tau.
$$
Local heat flux is defined as
$$
J_n=\frac{1}{N_y}\sum_{m=1}^{N_y}\langle j_{nm}(t)\rangle_t=
\frac{1}{N_y}\sum_{m=1}^{N_y}\lim_{t\rightarrow\infty}\frac1t\int_0^t j_{nm}(\tau)d\tau,
$$
where the energy flux from the node $(n,m)$ to the node $(n+1,m)$ is defined as
$$
j_{nm}=-V'(u_{n+1,m}-u_{n,m})(\dot{u}_{nm}+\dot{u}_{n+1,m})/2.
$$
\begin{figure}[tb]
\includegraphics[angle=0, width=0.9\linewidth]{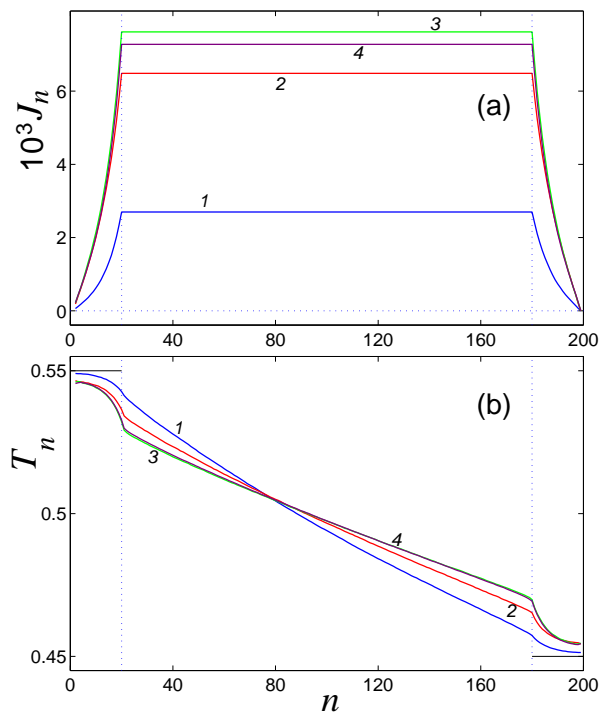}
\caption{
(Colour on-line)
Distribution of (a) local heat flux $J_n$ and (b) local temperature $T_n$
along the rectangular stripe of the rotators lattice with the length
$N_t+N_x+N_t=200$ (the length of the interior fragment is $N_x=160$, the length
of the ends inserted into thermostats is $N_t=20$, temperatures
of the thermostats $T_+=0.55$, $T_-=0.45$)
for width $N_y=1$, 2, 4 and 8 (curves 1, 2, 3 and 4).
}
\label{fig01}
\end{figure}

The local distribution of the average heat flux $J_n$
and the average temperature $T_n$ along the axis are presented in Fig. \ref{fig01}.
One observes that in the interior fragment of the stripe $N_t<n\le N_t+N_x$
constant energy flux $J_n\equiv J$ and linear temperature profile $T_n$  are formed.
It allows us to determine the heat conductivity coefficient
using only interior fragment of the stripe with width $N_y$
\begin{equation}
\kappa(N_x,N_y)=JN_x/(T_{N_t+1}-T_{N_t+N_x}). \label{f8}
\end{equation}
The value of the limit
\begin{equation}
\kappa(N_y)=\lim_{N_x\rightarrow\infty}\kappa(N_x,N_y) \label{f9}
\end{equation}
corresponds to the heat conductivity coefficient under average temperature
 $T=(T_++T_-)/2$.
\begin{figure}[tb]
\includegraphics[angle=0, width=0.875\linewidth]{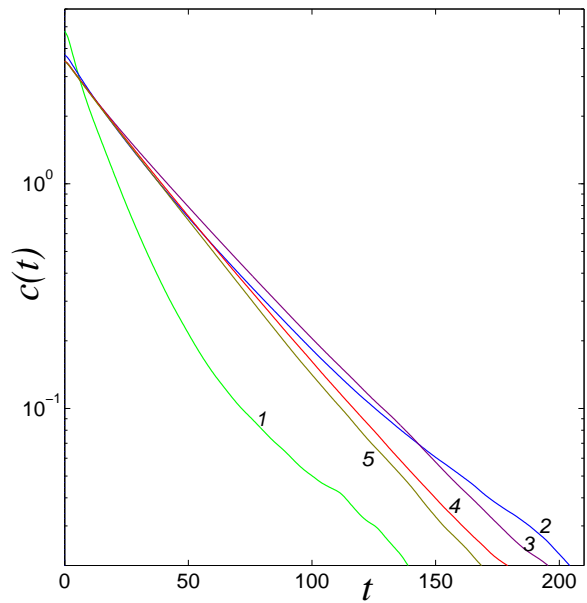}
\caption{
(Colour on-line)
Exponential decay of the autocorrelation
function $c(t)$ for stripe of the 2D sinh-Gordon lattice
 with width $N_y$=1,\,2,\,4,\,6 and 8 (curves 1,\,2,\,3,\,4 and 5).
The parameter of the substrate potential is\,$\omega_0=1$, and temperature~$T=6$.
}
\label{fig02}
\end{figure}

The heat conductivity coefficient may be also obtained
with the help of Green-Kubo relation \cite{Kubo}
\begin{equation}
\kappa(N_y)=\lim_{t\rightarrow\infty}\lim_{N_x\rightarrow\infty} \frac{1}{N_xN_yT^2}\int_0^tc(\tau)d\tau,
\label{f10}
\end{equation}
where $c(t)=\langle J_s(\tau)J_s(\tau-t)\rangle_\tau$ is an autocorrelation function
of the total heat flux in the chain $J_s(t)=\sum_{n=1}^{N_x}\sum_{m=1}^{N_y} j_{n,m}(t)$
with periodic boundary conditions.

In order to calculate the autocorrelation function $c(t)$ we
considered a cyclic stripe (torus) with length $N_x=2000$, 4000, 10000 entirely coupled
to the Langevin thermostat with temperature $T$. After achieving the thermal equilibrium
of the stripe with the thermostat, the system was detached from the thermostat and Hamiltonian
dynamics was simulated. To improve the accuracy, the results were averaged over $10^4$
realizations of the initial thermal distribution.

The heat conductivity coefficient of a 2D lattice in the axial direction
may be defined in terms of the limit of
the heat conductivity as the width of the stripe approaches infinity:
\begin{equation}
\kappa=\lim_{N_y\rightarrow\infty}\kappa(N_y). \label{f11}
\end{equation}
So, the question on the convergence of the heat conductivity of the
2D scalar lattice is formally reduced to the question of existence of the finite
limit (\ref{f11}). Fortunately, it turned out that one requires relatively small width
of the stripe (or perimeter of the cylinder cross-section) to achieve the convergence.

\section{Sinh-Gordon model}
Let us model the heat transport in a 2D lattice with a set of
potentials (\ref{f3}). Parameter of the substrate potential is adopted to be  $\omega_0=1$ and
the temperature  $T=6$.

Direct numerical modelling of the heat transport and analysis of the
behaviour of the autocorrelation function $c(t)$ show that a rectangular
stripe of the 2D lattice sinh-Gordon has finite heat conductivity for any stripe width.
The autocorrelation function for $t\rightarrow\infty$ approaches zero
exponentially -- see Fig. \ref{fig02}.
So, the integral of the Green-Kubo relation (\ref{f10})  converges in all considered cases.
\begin{figure}[t]
\includegraphics[angle=0, width=1\linewidth]{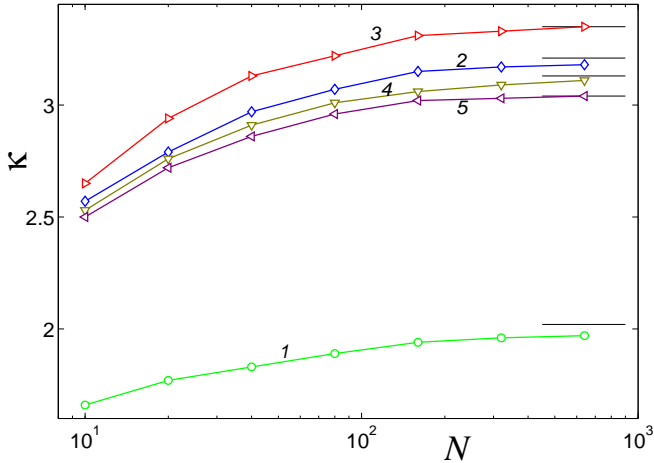}
\caption{
(Colour on-line)
Dependence of the heat conduction coefficient of the stripe of the sinh-Gordon
lattice $\kappa$ on the length of $N_x$ for width $N_y=1$,\,2,\,4,\,6,\,8 (curves 1,\,2,\,3,\,4,\,5).
Horizontal lines are calculated using Green-Kubo relation.
The parameter of the substrate potential is $\omega_0=1$, temperature~$T=6$.
}
\label{fig03}
\end{figure}

The dependence of the heat conductivity coefficient $\kappa(N_x,N_y)$ on length $N_x$
and width $N_y$ of the lattice are presented in Fig. \ref{fig03}. As it shown on the Figure,
the heat conductivity coefficient converges for any width of the stripe,
and the limit value of the heat conductivity practically ceases to
depend on width of the stripe for $N_y\ge 10$ -- see Table \ref{tab1}.
An increase of the width initially leads to an increase of the heat conductivity.
The maximal value of the heat conductivity is obtained for the width of the stripe $N_y=4$,
further increase of the width, however, leads to a decrease  of the heat conductivity with
further convergence. Limit values of the heat conduction coefficient for long strips correspond
to the values obtained from Green-Kubo relation. So, the two-dimensional scalar sinh-Gordon model,
similarly to its 1D counterpart,  has convergent heat conductivity.
\begin{largetable}
\caption{
Dependence of the heat conductivity coefficient of the rectangular
stripe $\kappa$ on the width $N_y$.
Sinh-Gordon model, parameter $\omega_0=1$, temperature $T=6$.}
\label{tab1}
\begin{tabular}{l|ccccccccccc}
\hline
\hline
$N_y$ & 1 & 2 & 3 & 4 & 5 & 6 & 7& 8 & 9 & 10 & 12 \\
\hline
$\kappa$ & 2.02 & 3.22 & 3.21 & 3.34 & 3.06 & 3.13 & 2.97 & 3.04 & 2.99 & 3.02 & 3.01\\
\hline
\hline
\end{tabular}
\end{largetable}

\section{Frenkel-Kontorova model}
The next set of potentials explored is 2D counterpart of  famous Frenkel-Kontorova (FK) chain (\ref{f4}).
We will use the value $\epsilon=5$  for a substrate parameter and examine the heat transport
at the temperature $T=5$. These values provide the fastest convergence of the heat conductivity
in the one-dimensional chain \cite{SG03}.
\begin{figure}[tb]
\includegraphics[angle=0, width=0.98\linewidth]{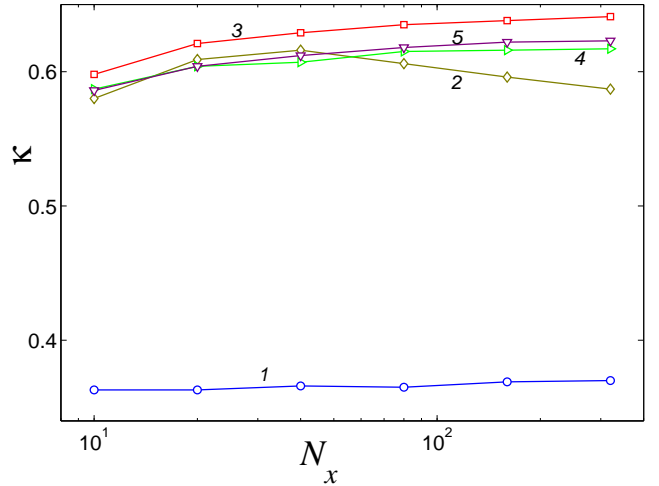}
\caption{
(Colour on-line)
Dependence of the heat conduction coefficient of the stripe of the FK  lattice $\kappa$
on the length of $N_x$ for width $N_y=1$,\,2,\,4,\,8,\,16 (curves 1,\,2,\,3,\,4,\,5).
The parameter of the substrate potential is $\epsilon=5$, temperature $T=5$.
}
\label{fig04}
\end{figure}

Direct numerical modelling of the heat transport and analysis of the
behaviour of the autocorrelation function show that a rectangular
stripe of the 2D scalar FK lattice also has finite heat
conductivity for any stripe width. The dependence of the heat conductivity coefficient
$\kappa(N_x,N_y)$ on the length $N_x$ and the width $N_y$ of the rectangular
section of the lattice is presented in Fig. \ref{fig04}. The heat conduction coefficient converges for
any width of the stripe, and the limit value of the heat conductivity practically ceases
to depend on width of the stripe for $N_y\ge 8$ -- see Table \ref{tab2}.

So, the two-dimensional scalar FK model as well as the one-dimensional model has finite heat
conductivity. Let us notice that in work \cite{barik07} the heat conductivity of the
2D FK lattice was considered for lower values of the parameter $\epsilon=1$.
For such a value of $\epsilon$ the heat conductivity of the 1D FK chain converges slowly, and
the rate of the convergence in the 2D lattice is even slower.
The work \cite{barik07} used lengths $N_x\le 250$, that is far below the length required
for the convergence, and creates wrong impression of infinite logarithmic growth of the heat conductivity
\begin{largetable}
\caption{
Dependence of the heat conductivity coefficient of the rectangular
stripe $\kappa$ on the width $N_y$.
FK model, parameter $\epsilon=5$, temperature $T=5$.
}
\label{tab2}
\begin{tabular}{l|cccccccccc}
\hline
\hline
$N_y$ & 1 & 2 & 3 & 4 & 5 & 6 & 8 & 12 & 16 \\
\hline
$\kappa$ & 0.344 & 0.668 & 0.650 & 0.642 & 0.620 & 0.618 & 0.620 & 0.618 & 0.615\\
\hline
\hline
\end{tabular}
\end{largetable}

\section{Rotators model}
Finally, we simulate the heat transport in the 2D lattice of rotators with the set of interaction
potentials (\ref{f5}). Let us consider the heat transport for four
characteristic temperature values $T=0.5$, 0.7, 1.0, 2.0.

\begin{figure}[tb]
\includegraphics[angle=0, width=1\linewidth]{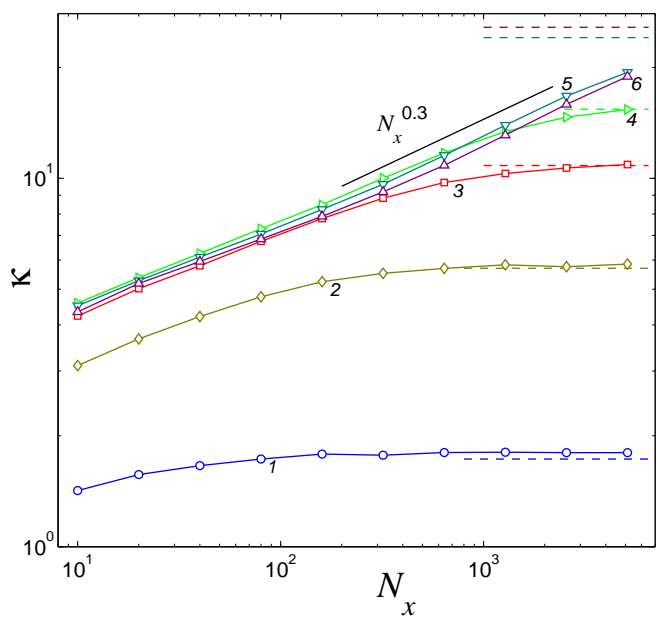}
\caption{
(Colour on-line)
Dependence of the heat conduction coefficient of the stripe of the rotators lattice $\kappa$
on the length of $N_x$ for width $N_y=1$, 2, 4, 6, 8 (curves 1, 2, 3, 4, 5). The temperature is $T=0.7$.
Horizontal lines are calculated using Green-Kubo
relation.
}
\label{fig05}
\end{figure}

Modelling of the heat transport for $T\ge 0.7$ clearly demonstrates
that a stripe with finite width always possesses a finite heat conductivity
coefficient that converges to a finite value as the width of the stripe
increases. The higher the temperature, the faster the convergence. For $T=2$ the maximal
heat conductivity is obtained for lengths $N_x\sim 10$ independently of the width of the stripe.
For  $T=1$ the convergence is established for lengths $N_x\sim 10^3$ (the maximal heat conductivity
refers to the stripe with width $N_y=4$),and for $T=0.7$ -- for lengths
$N_x\sim 10^4$ (the maximal heat conductivity
refers to the stripe with width $N_y=8$ -- see Fig. \ref{fig05})

The situation changes for temperature  $T=0.5$. Here the convergence in the chain of rotators
(width $N_y=1$) is obtained for length $N_x\sim 10^3$. An increase in the width of the stripe
leads to  slower pace of the convergence --  in the stripe with width $N_y=2$ the convergence
is now obtained for length $N_x\sim 10^4$. For increased width of the stripe the heat conductivity
$\kappa(N_x)$ grows on lengths $N_x\le 10^4$ as a power function $N^\alpha$ with the exponent $\alpha=0.3$.
However, it does not mean that the heat conductivity is infinite.
Such a behaviour of the heat conductivity takes place for length $N_x\le 10^3$ and temperature
$T=0.7$, but for longer lattices the heat conductivity converges -- see Fig. \ref{fig05}.

Mechanism for formation of the normal heat conductivity for temperatures $T=0.5$, 0.7 is
the same -- phonons scatter on discrete rotobreathers \cite{GS00}.
As the temperature decreases, the density of rotobreathers also decreases; thus one expects
the convergence of the heat conductivity at longer lattices. Analysis of the behaviour
of the autocorrelation functions shows that for low temperature $T=0.5$ the convergence of the
heat conductivity is expected at lengths $N_x\sim 10^5$, that are
still difficult for numerical simulations.
\begin{figure*}[tb]
\includegraphics[angle=0, width=0.81\linewidth]{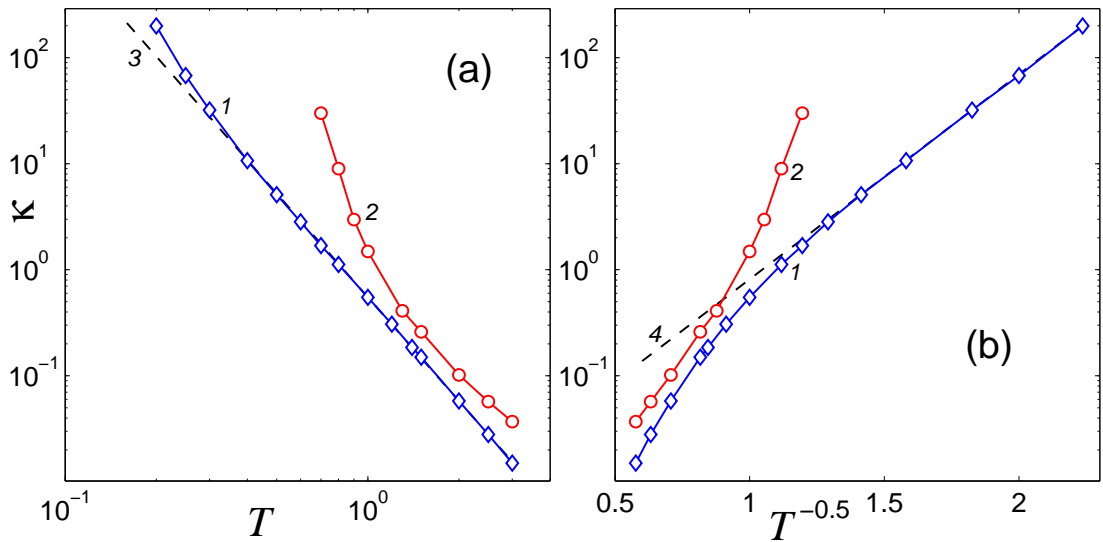}
\caption{
(Colour on-line)
Dependence of the heat conduction coefficient $\kappa$ on (a) temperature $T$ and (b)
on $1/\sqrt{T}$ for the chain (curve 1)
and the lattice of rotators (curve 2). The dashed line 3 corresponds to the relation
$\kappa=0.55T^{-3.25}$, line 4 -- to the relation
$\kappa=0.0095\exp(4.45/\sqrt{T})$.
}
\label{fig06}
\end{figure*}

To further explore the observed differences between the 1D and 2D rotator lattices, we consider the
dependence of the heat conduction coefficient on the temperature. This dependence  is presented
in Fig. \ref{fig06}.
One can see that the heat conductivity monotonically decreases with an increase in temperature.
A crossover between different types of asymptotic behavior in different temperature ranges
is observed. Figure \ref{fig06} (a) demonstrates a power law decrease ($\kappa=0.55T^{-3.25}$)
for temperatures $T\ge 0.4$ (similar behavior was also obtained in \cite{Li15}).
For $T\le 0.5$ the temperature dependence of the heat conduction coefficient is better approximated
by the relation $\kappa=0.0095\exp(4.45/\sqrt{T})$ -- see Fig. \ref{fig06} (b).

The heat conductivity of the 2D
lattice is significantly higher than that of the 1D chain. It is possible to conjecture that
two-dimensionality of the lattice allows phonons to bypass the localized rotobreathers,
with corresponding increase of their mean free path. This effect is strongly pronounced at low
temperatures -- for $T\rightarrow 0$ the heat conductivity of the 2D lattice grows much faster
than that of the one-dimensional chain.

\section{Conclusion}
The presented simulations of the heat transport in a two-dimensional
scalar lattices of rotators,  sinh-Gordon and FK  demonstrate that, as one would expect, if the
1D lattice possesses convergent heat conductivity, then the corresponding 2D lattice has the
same property. Different results obtained in earlier reports can be attributed to insufficient
size of the simulated system.

However, this finding does not imply that the 2D lattices with convergent heat conductivity behave
similarly to their 1D counterparts and do not introduce any new physics. First of all, it
turns out that for the convergence one requires large length only in one direction. Genuine 2D
behaviour reveals itself even for relatively narrow strips, thus drastically reducing the
computational efforts. The addition of the dimension can lead to deceleration of the convergence
-- in 2D lattice the convergence of the heat conductivity can require much longer linear size
than in the 1D chain. With a decrease of the temperature $T\searrow 0$ the heat conductivity
in 2D grows much faster than in the corresponding 1D chain. Besides, the convergence with
the width of the strip is not monotonous -- very narrow strips can have larger heat conductivity
than both the 1D chain and the 2D limit. Possible explanation is a competition of two factors
appearing in 2D. From one side, the phonons can bypass the localized excitations. From the other side,
additional dimensionality provides much more possibilities for the phonon-phonon scattering.
The former factor is very pronounced for the narrow strips, and the latter -- for wider strips,
thus explaining at qualitative level the observed non-monotonicity.

\acknowledgments
The authors are very grateful to Israel Science Foundation (grant 838/13) for financial support.
A.V.S. is grateful to the Joint Supercomputer Center of the Russian Academy of Sciences
for the use of computer facilities.

\end{document}